\documentclass[runningheads,a4paper]{llncs}

\usepackage{amssymb}
\usepackage{graphicx}
\usepackage{amsmath}

\usepackage{array}
\newcolumntype{C}[1]{>{\centering\arraybackslash}p{#1}}

\usepackage{url}
\urldef{\mailsa}\path|{binh, takasu}@nii.ac.jp|    
\newcommand{\keywords}[1]{\par\addvspace\baselineskip
\noindent\keywordname\enspace\ignorespaces#1}

\begin{document}

\mainmatter  

\title{Co-Factorization Model for Collaborative Filtering with Session-based Data}

\titlerunning{Lecture Notes in Computer Science}

%
%
\author{Binh Nguyen\inst{1}
\and Atsuhiro Takasu\inst{2}}
\authorrunning{Lecture Notes in Computer Science: Authors' Instructions}

\institute{SOKENDAI (The Graduate University for Advanced Studies), Shonan Village, Hayama, Kanagawa 240-0193 Japan\\
\and
National Institute of Informatics,
2-1-2 Hitotsubashi, Chiyoda, Tokyo 101-8430, Japan\\
\mailsa\\}

%
%

\toctitle{Lecture Notes in Computer Science}
\tocauthor{Authors' Instructions}
\maketitle

\begin{abstract}
Matrix factorization (MF) is a common method for collaborative filtering. MF represents user preferences and item attributes by latent factors. Despite that MF is a powerful method, it suffers from not be able to identifying strong associations of closely related items. In this work, we propose a method for matrix factorization that can reflect the localized relationships between strong related items into the latent representations of items. We do it by combine two worlds: MF for collaborative filtering and item2vec for item-embedding. The proposed method is able to exploit item-item relations. Our experiments on several datasets demonstrates a better performance with the previous work.
\keywords{Recommender System, Matrix Factorization, Collective Matrix Factorization, Item Embedding, Collaborative Filtering}
\end{abstract}

\section{Introduction}
Recommender System is a powerful system that support users to discover the information by suggesting them the items that they may interest. For example, Amazon recommends products to users based on their shopping histories. The technology behind recommender system is to describe users and items, and finding how to relate them.

Algorithms for Recommender Systems are often divided into two types: \textit{Content-based} algorithms and \textit{Collaborative Filtering-based} algorithms (\textit{CF-based} algorithms). \textit{Content-based} algorithms recommend items to an user based on the contents of items that the user has consumed. For example, in an online music website, if a user often listened rock music in the past, the system likely recommends this user a rock song. \textit{CF-based} algoritms, in other hand, recommend items to users based on their past behaviors (e.g., rating scores, clicks, purchases). A \textit{CF-based} algorithm does not require any information about the items. One This work focus on CF approach.

A popular method for latent factor models is Matrix Factorization (MF) which transforms users and items to latent vectors by factorizing the user-item matrix into two matrices. One for users' latent factor vectors and one for items' latent factor vectors.  Missing values in the matrix (e.g., rating score or implicit feedback such as clicks, visits) are approximated by the the inner product of corresponding latent vectors. These predicted missing values are used as the relevant of user to item for unseen interactions \cite{salakhutdinov2008a,hu2008collaborative,koren2008factorization}. The assumption behind the MF is that the user-item interactions are independent, i.e., a user consume an item independently, this consumption does not depend on other consumptions. However, this is not true for some situations. For example, an item that a user inserts to the shopping cart is strongly related to other items in the same shopping cart; a venue that a user visits strongly depend on the previous venues that he have visited as well as the future venues on his trajectory. Due to this assumption, MF is poor at capturing strong associations among closely related items.

The goal of this work is to make original MF be able to detect the relationships of related items and reflect them into the latent factors. We consider related items are items that frequently co-occur in the same sessions (e.g., products in the same sessions). The more frequently the two items appear in same sessions, the more related they are. We do it by factorizing two matrix: user-item matrix and item-item session-based co-occurrence matrix (item-item matrix) simultaneously. In this model, the role of user-item matrix factorization is similar to original collaborative filtering algorithm, while the role of item-item matrix factorization is for adjusting the latent vectors of items in order to reflect item-item relationship.

\section{Preliminary}
Collaborative filtering-based approaches is commonly based on the user-item matrix $\mathbf{R}\in \mathbb{R}^{N \times M}$, where $N$ and $M$ are number of users and items respectively. Each element $r_{ui}$ of $\mathbf{R}$ records the relevant score of user $u$ for item $i$. In explicit case, $r_{ui}$ would be the rating scores that $u$ gave to $i$, while in implicit case, $r_{ui}$ would be number of clicks or number of visit to a venues.
\subsection{Matrix Factorization and Collective Matrix Factorization}
\subsubsection{Matrix Factorization}
Given the user-item matrix $R\in \mathbb{R}^{N\times M}$, a popular approach to map users and items into latent vectors is matrix factorization (MF) with respect to it. MF factorizes $\mathbf{R}$ in to two matrices $X\in R^{K\times N}$ and $Y \in R^{K\times M}$ where $\mathbf{X}$ and $\mathbf{Y}$ are users' latent vectors and items' latent vectors respectively, where $K$ is the number of factors in the latent space. MF finds optimal $\mathbf{X}$ and $\mathbf{Y}$ by minimizing the following loss function:
\begin{equation}
    \mathbb{L}(\mathbf{R}, \mathbf{X}, \mathbf{Y})=||\mathbf{W}\odot(\mathbf{R}-\mathbf{X}^T\mathbf{Y})||^2_F+\lambda_u\sum_{u=1}^N{||\mathbf{X}_u||^2_F}+\lambda_v\sum_{i=1}^{M}{||\mathbf{Y}_i||^2_F}
\end{equation}
where $\odot$ is the element-wise product (Hadamard product) of matrices. $\mathbf{W}$ is a binary indication matrix whose entry $w_{ui}$ indicates if user $u$ has consumed item $i$ or not, i.e., $w_{ui}=1$ if $r_{ui}>0$, otherwise $w_{ui}=0$. $||.||_F$ is the Frobenius norm of either a matrix or a vector. This optimization can be viewed as least square problem and easily solved by Stochastic Gradient Descent algorithm as described in \cite{salakhutdinov2008a}.

The relevant score of user $u$ to unconsumed item $i$ is approximated as the inner product of user $u$'s latent vector and item $i$'s latent vector as follows.
\begin{equation*}
    r_{ui}=\mathbf{X}_u^T\mathbf{Y_i}
\end{equation*}
\subsubsection{Collective Matrix Factorization}
Collective Matrix Factorization (CMF) \cite{SinghG_kdd08} factorizes multiple matrices simultaneously in order to exploit the data from multiple sources.
\subsection{Item2Vec by Matrix Factorization}
Inspired by Word2Vec, a model for word embedding (i.e., representation of words by latent vectors), model for item embedding is developed \cite{conf/recsys/BarkanK16}. Word2Vec is a method that aims to learn word representation that capture the association between a words to its surrounding words. The authors of Item2Vec bring that idea to the world of recommender systems. The authors viewed a set or basket of items is equivalent to the sequence of words. In Item2Vec, each item $v_i$ is associated with two latent vectors $x_i\in \mathbf{R}^K$ and $y_i\in \mathbf{R}^K$ that corespond to the target and context representations respectively. For a given set of items $\{v_i\}^K_{i=1}\subset \textbf{V}$, Item2Vec learns the item representations by maximizing the following function.

\begin{equation}
    \label{item2vecop}
    \frac{1}{K}\sum_{i=1}^{K}\sum_{j\neq i}^K\log\left(\sigma(x_i^Ty_j)\prod_{k=1}^{M}\sigma(-x_i^Ty_k)\right)
\end{equation}

where $\sigma(x)=1/{1+\exp(-x)}$, $N$ is a parameter that determines the number of negative examples to be drawn for each positive example.

$x_i$ and $y_i$ ($i=1, \dots, M$) are estimated by using stochastic gradient ascent to the optimization problem in Eq.\ref{item2vecop}. $x_i$ is then used as the latent representation of item $v_i$.

The problem of item-embedding also can be interpreted as the implicit factorization of point-wise mutual information matrix (PMI) shifted by $\log k$ as described in \cite{levy2014neural}. First, the PMI matrix $\mathbf{P}$ is formed where each element $p_{ij}$ is the point-wise mutual information of item $v_i$ and $v_j$ as follows.

\begin{equation*}
    p_{ij}=\log\frac{\#(v_i, v_j).D}{\#(v_i).\#(v_j)}
\end{equation*}

where $\#(v_i, v_j)$ is the number of times that item $v_j$ appears in the context of item $v_i$. $\#(v_i)$ and $\#(v_j)$ are number of times $v_i$ and $v_j$ are consumed respectively. $D$ is number of item-context pairs.

Finally, the item-embedding is performed by factorizing the shifted positive matrix of PMI (SPPMI matrix):
\begin{equation*}
    \text{SPPMI}(i, j)=\max\{p_{ij}-\log k, 0\}
\end{equation*}

\section{Joint Factorization Model for Session-based Data}
This model integrate original MF and Item2Vec into a unified scheme. We do it by factorizing both user-item matrix and item-item matrix at the same time. This model is formulated as follows.

\begin{multline}
    L(R, V, X, Y, Z) = \sum_{i=1}^N\sum_{j=1}^M {I_{ui}(R_{uj}-X_u^TY_i)^2}+\sum_{i=1}^M\sum_{j=1}^M{(V_{ij}-Y_i^TZ_j)^2}\\+\lambda_x\sum_{i=1}^N{||X_u||^2_F}+\lambda_y\sum_{j=1}^{M}{||Y_i||^2_F}+\lambda_z\sum_{j=1}^{M}{||Z_j||^2_F}\\
    s.t. \text{ } X_u, Y_i, Z_j>0 \text{ }(u=1, \dots, N; i,j=1,\dots, M)
\end{multline}

Where $I_{ui}$ is to indicate if user $u$ has consumed item $i$ or not. It takes value 1 if use $u$ has consumed item $i$, otherwise, it takes value 0. The meaning of $I_{ui}$ is to skip non-interact $user-item$ pairs from the optimization process.

\section{Experimental Study}
Our goal of the experiments is to evaluate the effect of incorporating the localized item-item relationships into the model. In this section, we first describe the settings of experiments and then show the experimental results.
\subsection{Datasets}
We evalutate our recommedation method using two public datasets.
\subsubsection{Global-scale Check-in Dataset of Foursquare (Foursquare Global)} \cite{journals/tist/YangZQ16,journals/jnca/YangZCQ15}: This dataset includes long-term (about 18 months from April 2012 to September 2013) global-scale check-in data collected from Foursquare. It contains 33,278,683 checkins by 266,909 users on 3,680,126 venues (in 415 cities in 77 countries). Those 415 cities are the most checked 415 cities by Foursquare users in the world, each of which contains at least 10K check-ins.
    
Foursquare Global is a timestamped dataset. Each check-in information is provided in the form of a tuple: ($user\_id, venue\_id, timestamp$).
    
From this dataset, we construct the user-item matrix by binarizing the visit history data (multiple visit to a venue are considered as single visit). In other words, the user-item matrix is a binary matrix $R$ where $R_{ui}=1$ if user $u$ has visited venue $i$ at least one time, otherwise, $R_{ui}=0$.
    
We construct the item-item matrix by creating the SPPMI matrix of session-based co-occurrence. In this experiments, we consider each session as the set of check-ins where the time interval between two successive check-ins are less than 6 hours. After that, we create the session-based co-occurrence matrix which each entry is the number of times two venues $i$ and $j$ appear in same sessions. Finally, the SPPMI is calculated from the session-based co-occurrence matrix.
\subsubsection{Last.fm 1K users dataset (Last.fm 1K)} \cite{books/daglib/0025137}: The dataset contains (\textit{user, timestamp, artist, song}) tuples collected from Last.fm API. This dataset represents the whole listening habits (till May, 5th 2009) for nearly 1,000 users. We construct the user-item matrix and item-item matrix by the same way as we do with Foursquare Global dataset above.

For each dataset, we divide into 3 sets: training/validation/test. We randomly take $80\%$ as training dataset for learning the parameters of the model. The remaining $20\%$ is used as ground truth for testing.

\subsection{Evaluation Metrics}
We use the top-$k$ method for recommendation. The learned model calculates the relevant score of each unconsumed item for each user and then ranks the venue based on the predicted scores. Top-$k$ items that have largest relevant scores of each user will the user as the recommendation list. We use $Recall$@k and Precision@$k$ as metrics for evaluating evaluating the quality of the recommendations. Recall@$k$ for each user show what percentage of the consumed items appear in the top-$k$ list. Precision@$k$ indicates what percentage of the recommendation list appears in the list of consumed items. Both Recall@$k$ and Precision@$k$ does not care about the order of items in the recommendation list.

If we define $S_u(k)$ and $V_u$ are the top-$k$ recommendation list and consumed items of user $u$ respectively, Recall@$k$ and Precision@$k$ are defined as follows.
\begin{equation}
    \begin{aligned}
    Recall@k &= \frac{1}{N}\sum_{u=1}^N\frac{|S_u(k) \cap V_u|}{|V_u|}\\
    Precision@k &= \frac{1}{N}\sum_{u=1}^N\frac{|S_u(k) \cap V_u|}{k}
    \end{aligned}
\end{equation}

\subsection{Competing Methods}
We compare the SessionCoFactor with following competing methods.
\subsubsection{Cofactor}: Cofactor \cite{confrecsysLiangACB16} factorizes user-item and item-item matrices simultaneously as we do in this research. The difference of our method with Cofactor lie in the construction of the item-item matrices. In CoFactor, the item-item matrix is constructed based on the co-occurrence of items in the same user's consumption list, while our item-item matrix is constructed based on the session-based co-occurrence matrix in order to exploit the associations of related items. 

\subsubsection{Weighted Matrix Factorization (WMF)}  \cite{hu2008collaborative}: a general method for matrix factorization in which each element $r_{ui}$ of user-item matrix is associated with a confident weight $w_{ui}$
\subsection{Experimental Results}
We learn the model with number of factors are 10, 20, 30 since these choices resulted in the best model performance for all our model as well as competing methods. Result for Recall@$k$, nDCG@$k$ and MAP@$k$ for Foursquare Global data are shown in Table \ref{result_foursquare}.

\begin{table}[ht]
\centering
\begin{tabular}{C{3cm} C{2cm} C{2cm} C{4cm}}
\hline
& WMF & CoFactor & Session-based CMF\\
\hline
    Recall@20            & 0.1492 & 0.1563 & \textbf{0.1612}  \\
Recall@50           & 0.2336 & 0.2350 & \textbf{0.2427}  \\
nDCG@20     & 0.1359 & 0.1608 & \textbf{0.1608}  \\
nDCG@50     & 0.1713 & 0.1928 & \textbf{0.1942}  \\
\hline
\label{result_foursquare}
\end{tabular}
\caption{Comparisonal results between Session-based CMF, CoFactor \cite{confrecsysLiangACB16} and WMF \cite{hu2008collaborative}. Session-based CMF performs better results accross all metrics}
\end{table}

\section{Related Work}
Many matrix factorization based methods are developed for collaborative filtering such as in \cite{salakhutdinov2008a,hu2008collaborative,pan:icdm08,nguyen2015city}. Hu et al. \cite{hu2008collaborative} propose weighted matrix factorization for user-item implicit feedback matrix. They treat the user-item matrix as a binary matrix and associate each element of the matrix with a confident weight which identify the confident of the element. The more frequency the interaction of a user to an item is observed, the higher confident weight is. They then formulate the factorization as the optimization problem that minimizes the squared error over the training set. Salakhudinov et al. \cite{salakhutdinov2008a} assume the elements of a user-item rating matrix given user and item latent factor vectors as a Gaussian distribution. The latent factor vectors are found by maximizing the log-likelihood. Goplan et al. \cite{journals/corr/GopalanHB13} introduced a Poisson-distribution based factorization model that factorizes user-item matrix.

The common point of the above methods for matrix factorization is that they assume that the user-item interactions are independent, thus can not capture the relationships between strong related items into the latent representations.

Collective Matrix Factorization \cite{SinghG_kdd08} proposed a framework for factorizing multiple related matrices simultaneously, in order to exploit information from multiple sources. For example, if the item-genre matrix exists, one can factorize both user-item and item-genre matrices in a share latent space. This approach is able to incorporate the side information (e.g., genre information of items) into the latent factor model.

CoFactor \cite{confrecsysLiangACB16} is a model that based on the Collective Matrix Factorization \cite{SinghG_kdd08}. It factorizes user-item matrix and item-item matrix at the same time in a shared latent space. There contribution lie in the construction of item-item matrix, which relied on the co-occurrence of items in the consumed list of users.

The main different of our method with Collective Matrix Factorization \cite{SinghG_kdd08} and CoFactor \cite{confrecsysLiangACB16} is how we build the item-item matrix in order to pull out the strong related items which frequently co-occur in same sessions.

Item2Vec \cite{conf/recsys/BarkanK16} is a method that bring the idea Word2Vec \cite{levy2014neural}, a word-embedding method, in to the world of recommender system. This method transforms items into a latent spaces (item embedding) in which items that frequently appear in same contexts should be located close in the space. This method is able to capture the relations between closely related items. Our method inspires from this research, however, we just user the item-item relations as a mean to adjust the latent factor vectors learnt by the factorization of user-item matrix.

\section{Discussion and Future Work}
We examined the effect of session-related items to the performance of recommendation. We proposed a method that combine the power of two worlds: collaborative filtering by MF and item-embedding with item-context is the items in same sessions. Our goal is to propose a latent factor model that refelect the strong associations of closed related items into the latent representations. The empirical results showed that our method performed better results compare with weighted matrix factorization \cite{hu2008collaborative} and CoFactor \cite{confrecsysLiangACB16} in across the metrics. This results show that the associations between related items take an important role in the quality of the recommendation list.

There are several directions to extend this work. It would be interesting to learn the model to optimize the ranking instead of the least square problem, especially in the case of POI recommendation.

\bibliographystyle{unsrt}
\bibliography{mybib}
\end{document}